\documentclass[12pt]{article}
\usepackage{eqsection,latexsym,epsf}
\usepackage{amsfonts}
\usepackage{epsfig}
\begin{document}

\title{The statistical mechanics of turbo codes.}

\author{
  \\
  {\small Andrea Montanari}              \\[-0.2cm]
  {\small\it Scuola Normale Superiore and INFN -- Sezione di Pisa}  \\[-0.2cm]
 {\small\it I-56100 Pisa, ITALIA}          \\[-0.2cm]
  {\small Internet: {\tt  montanar@cibs.sns.it}}   
  \\[-0.2cm]
  \\[-0.1cm]  \and
  {\small Nicolas Sourlas}              \\[-0.2cm]
 {\small\it Laboratoire de Physique Th{\'e}orique de l' Ecole Normale
Sup{\'e}rieure  
\footnote {UMR 8549, Unit{\'e}   Mixte de Recherche du 
Centre National de la Recherche Scientifique et de 
l' Ecole Normale Sup{\'e}rieure. } }         \\[-0.2cm]
  {\small\it 24 rue Lhomond, 75231 Paris CEDEX 05, France.}          \\[-0.2cm]
  {\small Internet: {\tt sourlas@lpt.ens.fr}}   \\[-0.2cm]
  {\protect\makebox[5in]{\quad}}  % To force authors' names to be written
                                  %   vertically, one above another.
                                  % (\author seems to put them side-by-side
                                 %   if there is room.)
 \\
}
\vspace{0.5cm}

\def\tg{\mbox{\boldmath $\tau$}}
\def\sg{\mbox{\boldmath $\sigma$}}
\def\thg{\mbox{\boldmath $\theta$}}
\def\Gg{\mbox{\boldmath $\Gamma$}}
\def\Hg{\mbox{\boldmath $H$}}
\def\Jg{\mbox{\boldmath $J$}}
\def\Cg{\mbox{\boldmath $C$}}
\def\Bg{\mbox{\boldmath $B$}}
\def\zb{\mbox{\boldmath $0$}}
\def\<{\langle}
\def\>{\rangle}
\def\arctanh{\mbox{arctanh}}

\maketitle
\thispagestyle{empty}   % Suppress page number on front page.

\vspace{0.2cm}

\begin{abstract}
The ``turbo codes'', recently proposed by Berrou
et. al. \cite{PrimoBerrou} 
are written as a disordered spin Hamiltonian. It is shown  
that there is a threshold $\Theta$ such that 
for signal to noise ratios $v^2 / w^2 > \Theta$ the error probability per 
bit vanishes in the thermodynamic limit, i.e. the limit of infinitly 
long sequences. The value of the threshold has been computed for 
two particular turbo codes. It is found that it depends on the code. These 
results are compared with numerical simulations. 
\end{abstract}

\vspace{4.2cm}
\begin{flushleft} 
LPTENS 99/29    
\end{flushleft}

\clearpage

\section{Introduction.}

The recent invention of ``turbo codes'' by Berrou and Glavieux 
\cite{PrimoBerrou} is considered a major  breakthrough in communications. 
For the first time one can communicate almost error-free for signal to noise 
ratios very close to the theoretical bounds of information theory.
Turbo codes are fastly becoming the new standard for 
error correcting codes in digital communications. 
The invention of turbo codes and their iterative 
decoding algorithm was empirical. 
There is no theoretical understanding of why they are so successfull. 
The decoding algorithm is thought to be an approximate algorithm. 
We think that turbo codes are interesting, even outside the context 
of communication theory, because they provide a non trivial example of 
a disordered system which can be studied numerically with a fast algorithm.\\ 
In this paper we will study turbo codes and turbo decoding
 using the modern tools 
of statistical mechanics of disordered systems. 
One of us has already shown in the 
past \cite{Sourlas1} that there is a mathematical equivalence between 
error correcting codes 
and theoretical models of spin glasses. In particular the logarithm of the 
probability for any  given signal, 
conditional  on the communication channel output, 
has the form of a spin glass Hamiltonian. We will construct the  Hamiltonian 
which corresponds to the turbo codes and study its properties. 
This will clarify why they are so successfull. 
In particular we will show that there is a threshold $\Theta$ such that 
for signal to noise ratios $v^2 / w^2 > \Theta$ the average 
error probability per bit $ \overline{P_e} $ 
vanishes in the thermodynamic limit, i.e. the limit of infinitly 
long sequences. In $ \overline{P_e} $ the average is taken over a large 
class of turbo codes (see later) and over ``channel'' noise.
The rate of these codes is finite. 
The value of the threshold has been computed for 
 two particular turbo codes. It was found that it depends on the code. 
We also compare these results with numerical simulations.\\ 
Our results are typical of 
the statistical mechanics approach: we study only the average performance 
 of turbo codes, not the performance of any particular one. 
Furthermore there exist ``very few'' particular codes
 performing ``much worse'' than the average.\\
Let us first briefly remind the connection between error-correction codes 
and spin-glass models. 
 In the  mathematical theory of communication   
both the production of information and its transmission 
 are considered as probabilistic events. A source is producing 
information messages according to a certain 
probability distribution. Messages of length $N$ are sequencies of $N$ 
symbols or ``letters of an alphabet'' $a_1,a_2, \cdots, a_N$. 
We will assume for simplicity  a binary alphabet, i.e. $a_i = 0$ or $1$ and 
that all symbols are equally probable. Instead of $a_i$ we can 
equally well use Ising spins 
\begin{eqnarray}
\sigma_i = (-1)^{a_i} = \pm     1 
\end{eqnarray} 
 The messages are sent through 
a noisy transmission channel. 
 If a $ \sigma  = \pm     1 $ is sent through the transmission 
channel, because of the noise, the output will be a real number 
$ \sigma ^{\mbox{out}}$, in general different from $ \sigma $. 
Again, the statistical properties 
of the transmission channel are supposed to be known. Let us call 
$ Q ( \sg^{\mbox{out}} | \sg ) d\sg ^{\mbox{out}} $ 
the probability for the transmission 
channel's output to be between $ \sg^{\mbox{out}} $ and 
$ \sg^{\mbox{out}}+ d\sg^{\mbox{out}}$,
when the input was $ \sg $. $ Q ( \sg^{\mbox{out}} | \sg )$
is supposed to be known.
For reasons of simplicity, we assume that 
the noise is independent for any pair of bits (``memoryless channel''),
i.e.
\begin{eqnarray} 
 Q ( \sg^{\mbox{out}} | \sg ) = 
\prod_{i} Q ( \sigma ^{\mbox{out}}_i | \sigma_i ) 
\end{eqnarray}
In the case of a memoryless channel and a gaussian noise:
\begin{eqnarray}
Q_{\mbox{\small{gauss}}} ( \sigma ^{\mbox{out}} | \sigma )
\equiv\frac{1}{\sqrt{2\pi w^2}}
\exp\left\{-\frac{(\sigma ^{\mbox{out}}-\sigma)^2}{2w^2}\right\}
\label{GaussianChannel}
\end{eqnarray}
Shannon 
calculated the channels capacity $ {\cal C }$, i.e. the maximum 
information per use 
of the channel that can be transmitted. 
\begin{eqnarray}
 {\cal C }_{\mbox{\small{gauss}}} = 
\frac{1}{2} \log_{2} (1 + \frac{v^2}{w^2} ) 
\label{GaussianCapacity}
\end{eqnarray}
where $ v^2 $ is the signal power.\\ 
Under the above assumptions, communication 
is a statistical inference problem. Given the transmission 
channel's output and the statistical 
properties of the source and of the channel, 
 one has to infer what message was sent. In order 
to reduce communication errors, 
one may introduce (deterministic) redundancy into the 
message (``channel encoding'') and use this redundancy to 
infer the message sent through the channel (``decoding'').
The algorithms which transform the source outputs to 
redundant messages are called error-correcting codes. 
More precisely, instead of sending the $N$ original bits 
$ \sigma_{i}  $, one sends $M$ bits $J_{k}^{\mbox{in}} $, 
$ k =1, \cdots , M $,$M > N $, 
constructed in the following way
\begin{eqnarray}
 J_{k}^{\mbox{in}} \ = \ C^{(k)}_{i_1...i_{l_k}} \ \sigma_{i_1} \cdots 
\sigma_{i_{l_k}} 
\end{eqnarray}
where the ``connectivity'' matrix 
$ C^{(k)}_{i_1\dots i_{l_k}} $ has elements zero or one. 
For any $k$, all the $ C^{(k)}_{i_1\dots i_{l_{k}}} $ except from one 
are equal to zero, i.e. the $J_{k}^{\mbox{in}} $ are equal to $ \pm     1$.
 $ C^{(k)}_{i_1\dots i_{l_k}} $ defines the code, 
i.e. it tells from which 
of the $\sigma$'s to construct the $k$th bit of the code.\\
This kind of codes 
are called parity checking codes because $J_{k}^{\mbox{in}} $ counts the 
parity of the minusis among the $ l_k $ $\sigma$'s. 
The ratio $ R = N / M $ which specifies the redudancy of the code, 
is called the rate of the code.\\
Knowing the source probability, the noise  probability, the code and 
the channel output, one has to infer the message that was sent.
The quality of inference depends on the choice of the code.\\
According to the famous Shannon's channel encoding theorem, there exist 
codes which, in the limit of infinitly long messages, 
 allow error-free communication, provided the rate of the 
code $ R $ is less than the channel capacity $ {\cal C } $. This 
theorem says that such ``ideal'' codes exist, but does not say 
how to construct them.\\
We have shown that there exists a close mathematical relationship 
between error-correcting codes and theoretical models of disosdered 
systems. 
As we previously said, the output of the channel is a sequence of $M$ 
 real numbers 
$ \Jg^{\mbox{out}} = \{ J_{k}^{\mbox{out}}, \ k=1, \cdots , M \} $, 
which are random variables, 
obeying the probability distribution 
$ Q(J_{k}^{\mbox{out}} | J_{k}^{\mbox{in}} ) $.
 Once the channel output $ \Jg^{\mbox{out}}$ is known, 
it is possible to compute the probability $P( \tg | \Jg^{\mbox{out}} ) $ 
for any particular sequence 
$ \tg \ = \{ \tau_{i} , \ i=1, \cdots, N \} $ to be the {\it source } 
output (i.e. the information message).\\
More precisely, the equivalence between spin-glass models and 
error correcting codes is based on the following property.\\
The probability $P( \tg | \Jg^{\mbox{out}} ) $ for any sequence 
$ \tg $ to be the information message, 
conditional on the channel output $ \Jg^{\mbox{out}} $ is given by  
\begin{eqnarray}
 \ln P( \tg | \Jg^{\mbox{out}} ) \  = \ \mbox{const} \   +  \sum_{k=1}^{M} 
C^{(k)}_{i_1...i_{l_{k}}} \ B_{k} \ \tau_{i_1} \cdots 
\tau_{i_{l_{k}}}  \ \equiv \  - H(\tg )
\label{DistribuzionePerCodiceGenerico}  
\end{eqnarray}
where 
\begin{eqnarray}
B_{k} \equiv B ( J_{k}^{\mbox{out}})  \equiv
\frac{1}{2}  \ln \frac{Q(J_{k}^{\mbox{out}}| 1 )}{Q(J_{k}^{\mbox{out}} | -1 )} 
\label{DefinizioneB}
\end{eqnarray}
We recognize in this expression the Hamiltonian of a p-spin spin-glass 
Hamiltonian. The distribution of the couplings is determined 
by the probability $Q(J^{\mbox{out}} | J^{\mbox{in}} )$.\\
In the case when 
$ Q(J^{\mbox{out}} | J^{\mbox{in}} ) \ = \ 
Q(-J^{\mbox{out}} | - J^{\mbox{in}} )   $
 (the case of a ``symmetric channel''), $ B ( J^{\mbox{out}}) = 
 - B_{k} ( - J^{\mbox{out}}) $ and one recovers the  invariance of the 
spin-glass Hamiltonian under gauge transformations.\\ 
``Minimum error probability decoding'' (or MED), 
which is widely used 
in communications \cite{AlgoritmoViterbi}, 
consists in choosing the most probable 
sequence $ \tg^0 $. This is equivalent 
to finding the ground state of the above spin-glass Hamiltonian.\\
Instead of considering the most probable instance, 
one may only be interested in 
the most probable value  $\tau_{i}^{\mbox{\tiny{MAP}}} $  of the ``bit'' $\tau_i $ (Maximum A posteriori Probability or MAP decoding) \cite{AlgoritmoBCJR} 
which can be expressed in terms of the magnetization at 
temperature $T=1 / \beta $ equal to one \cite{Rujan}:
\begin{eqnarray}
\tau_{i}^{\mbox{\tiny{MAP}}}  = {\rm sign} \ ( m_i ) \quad ;\quad  m_i  \ = 
\frac{1}{Z} \ \sum_{ \tg } 
 \tau_{i} \ \exp\{-  H (\tg)\}  
\end{eqnarray}
where $H(\tg)$ is defined by Eq.(\ref{DistribuzionePerCodiceGenerico}).\\
It is remarkable that $ \beta = 1 $ coincides with the Nishimori 
temperature in spin glasses \cite{Nishimori}. 
MAP decoding is an essential ingredient in turbo decoding (see later).\\
When all messages are equally probable and the transmission channel is 
memoryless and symmetric, the error 
probability is the same for all input sequences. It is enough to compute 
it in the case where all input bits are equal to one. In this case, 
the error probability per bit $P_{e} $ is 
\begin{eqnarray}
 P_{e} \ = \frac{1}{2}(1-m^{(d)})\equiv 
\frac{1}{2}\left(1-\frac{1}{N} \sum_{i=1}^{N} \tau_{i}^{(d)}\right)
\end{eqnarray} 
and  $ \tau_{i}^{(d)} $ is the symbol sequence produced by the
decoding procedure. 
One can derive from this 
a very general lower bound for $P_{e} $, using the analog of the low 
temperature expansion. An obvious bound (for zero temperature
decoding) is provided 
by the probability $P^{(1)}_{e} $ that only one bit is 
incorrect, i.e. $\tau_j = -1 $ while all other bits are 
correct, i.e. $\tau_i = 1 $ for all $i \ne j$:
\begin{eqnarray} 
P_{e} \ge P^{(1)}_{e} = 
\mbox{Probability of } \left\{ \sum_{k\in\Omega(j)} B_{k} < 0 
\right\}
\label{ConnectivityBound}
\end{eqnarray}
where the $ \Omega(j) $ denotes the set of the couplings in which 
$\tau_j$ appears.\\
A necessary condition for transmitting without errors is that 
$\sum_{k\in\Omega(j)} B_k > 0  $ with probability one. 
This is only possible if  every spin appears in an infinite number 
of terms in the Hamiltonian. 
Let $l_k $ be the number of spins coupled through the coupling 
$B_{k}$.\\
The total number of spins beeing $N$, 
a spin appears on the average in 
\begin{eqnarray}
\frac{1}{N} \sum_{k=1}^{M} l_k = 
\frac{M}{N}\frac{1}{M} \sum_{k=1}^{M} l_k = \frac{{\bar l}}{R} 
\end{eqnarray} 
terms, where $ { \bar l } $  is the average of $l_k$ (the number of spins
 coupled together)   
and $ R $ is the rate of the code.\\
So a necessary condition  for a finite rate  code to achieve 
zero error probability, is that the average number of spins coupled
together diverges in the thermodynamic limit ($ N \to \infty $). 
This condition is realised in Derrida's random energy model \cite{DerridaREM} 
 which has been shown 
to be an ideal code \cite{Sourlas1} ( in that case $ R= 0 $ ).\\
We will show in the following 
that this is also true for the case of recursive turbo codes, 
while it is not true for non recursive turbo codes.
%
%*************************************************************************
%
\section{Convolutional codes.}
\label{ConvolutionalSection}
Convolutional codes are the building blocks of turbo codes. 
In this section we shall describe both non recursive and recursive 
convolutional codes and the corresponding spin models. The information message
 i.e. the source output (before encoding) will be denoted by:
\begin{eqnarray}
\tg  &\equiv & (\tau_1,\dots,\tau_N)
\end{eqnarray} 
It is convenient to think of the source producing a symbol per unit 
time, i.e. in $\tau_i $, $i$ denotes the time. 
For simplicity we consider a code of rate $R = 1/2$. 
The encoded message has the form:
\begin{eqnarray}
\Jg\equiv(J^{(1)}_1,\dots,J^{(1)}_N;J^{(2)}_1,\dots,J^{(2)}_N)
\end{eqnarray}
Any hardware implementation of a convolutional encoder contains a sequence
of $r$ memory registers. We shall call $r$ the range of the code.\\
Let's denote by $\Sigma_1(t),\dots,\Sigma_r(t)$  
the content of the memory registers at time $t$. 
At each time step the content of each memory register is
shifted to the right:
\begin{eqnarray}
\Sigma_{j+1}(t+1) = \Sigma_j(t) \quad\mbox{for}\quad j= 1,\dots,r-1
\end{eqnarray}
Moreover for convenience of notation we  define
\begin{eqnarray}
\Sigma_0(t) \equiv \Sigma_1(t+1)
\end{eqnarray}
and the following sequence of bits which we shall call the register
sequence:
\begin{eqnarray}
\sg  \equiv  (\sigma_1,\dots,\sigma_N), \quad 
\sigma_i  \equiv  \Sigma_0(i)
\end{eqnarray}
The sequence of the $\sigma$'s is a function of the source sequence (which 
may depend on the code):
\begin{eqnarray}
\tg & \mapsto & \sg(\tg)\equiv (\sigma_1(\tg),\dots,\sigma_N(\tg))
\label{register}
\end{eqnarray}
When not ambiguous we shall omit in the following the functional
dependence of $\sg$ upon $\tg$.\\
For non recursive convolutional codes this application is extremely simple: 
\begin{eqnarray}
\sigma_i(\tg) = \Sigma_0(i) = \tau_i
\label{RegistroNonRicorsivo}
\end{eqnarray}
The encoded message $\Jg$ is easily defined in
terms of the content of the register sequence:
\begin{eqnarray}
J^{(n)}_i = \prod_{j=0}^r (\Sigma_j(i))^{\kappa(j;n)} = 
\prod_{j=0}^r (\sigma_{i-j})^{\kappa(j;n)}
\label{encoding}\\
i = 1,\dots,N \ ; \ n = 1, 2\nonumber\\
\kappa(j;n)\in \{0,1\}\nonumber
\end{eqnarray}
We shall assume hereafter that $\kappa(0;1) = \kappa(0;2) = 1$.\\
To avoid redundancy we choose $r$ such that either $\kappa(r;1)$ or 
$\kappa(r;2)$ are different from $0$.\\ 
To make Eq.(\ref{encoding}) meaningful for $i = 1,\dots,r$ we define 
$\sigma_j = +1$ for $j\le 0$. Notice however that the exact definition
of $J^{(n)}_1,\dots,J^{(n)}_r$ is irrelevant in the thermodynamic limit.\\
The numbers $\kappa(j;n)$ define the code. Several conventions are
used  to give them in a compact form. A simple and useful one is the
following. To each of the two sets of numbers $\kappa(j;1)$ and $\kappa(j;2)$  
 is associated a polynomial on $\mathbb{Z}_2$ 
\begin{eqnarray}
g_n(x)=\sum_{j=0}^r \kappa(j;n)x^j.
\end{eqnarray}
The $g_n$ are called generating polynomials. In the same way we can 
associate a polynomial to the register sequence 
($ G(x) \equiv \sum_{j=1}^N b_j x^j \ ; \ \sigma_j \equiv (-1)^{b_j} $), 
to the source message
($ H(x) \equiv \sum_{j=1}^N a_j x^j \ ; \ \tau_j \equiv (-1)^{a_j} $)  
and to each part of the encoded message 
($ {\cal G}^{(n)} (x) \equiv \sum_{j=1}^N d_{j}^{(n)} x^j \ ; \ 
J^{(n)}_j \equiv (-1)^{d^{(n)}_j}$). With these definitions it is
evident that the correspondence (\ref{RegistroNonRicorsivo}) between 
the source and the register sequences for a non recursive
convolutional code implies:
\begin{eqnarray}
G(x) = H(x)
\label{RegistroNonRicorsivoPol}
\end{eqnarray}
and the encoding rule (\ref{encoding}) becomes
\begin{eqnarray}
{\cal G}^{(n)} (x) = g_n (x) G(x) = g_n(x)H(x) 
\label{nonrecursivepolyn}
\end{eqnarray}
A few examples  are the following:
\renewcommand{\theenumi}{(\alph{enumi})}
\begin{enumerate}
\item The simplest non trivial convolutional code has range $1$:
\begin{eqnarray}
J^{(1)}_i  =  \sigma_i\sigma_{i-1}&\Rightarrow& g_1(x) = 1+x
\label{CodiceSemplice1}\\
J^{(2)}_i  =  \sigma_i&\Rightarrow& g_2(x) = 1
\label{CodiceSemplice2}
\end{eqnarray} 
\label{CodiceSemplice}
\item A simple code with range $r=2$ whose 
behavior will be examined in what follows:
\begin{eqnarray}
J^{(1)}_i  =  \sigma_i\sigma_{i-1}\sigma_{i-2}
&\Rightarrow& g_1(x) = 1+x+x^2
\label{CodiceTipico1}\\
J^{(2)}_i  =  \sigma_i\sigma_{i-2}&\Rightarrow& g_2(x) = 1+x^2
\label{CodiceTipico2}
\end{eqnarray}
\label{CodiceTipico}
\item The code with range $r=4$ used by Berrou and collaborators 
to build the first example of turbo code:
\begin{eqnarray}
J^{(1)}_i  = 
\sigma_i\sigma_{i-1}\sigma_{i-2}\sigma_{i-3}\sigma_{i-4}
&\Rightarrow & g_1(x) = 1+x+x^2+x^3+x^4
\label{CodiceComplicato1}\\
J^{(2)}_i  =  \sigma_i\sigma_{i-4}&\Rightarrow & g_2(x) = 1+x^4
\label{CodiceComplicato2}
\end{eqnarray}
\label{CodiceComplicato}
\end{enumerate}
Recursive convolutional codes are most easily defined in terms of the 
generating polynomials. The difference with non recursive codes 
 is in the relation between the source 
 and the register sequences. In the non recursive case it was
given by Eq.(\ref{RegistroNonRicorsivo}) or by 
Eq.(\ref{RegistroNonRicorsivoPol}). In the recursive case one has:
\begin{eqnarray}
G(x) = \frac{1}{g_1(x)}H(x)
\end{eqnarray}
so that Eq. (\ref{nonrecursivepolyn}) gives 
\begin{eqnarray}
{\cal G}^{(1)}(x) = g_1(x)G(x) = H(x) \ , 
\quad {\cal G}^{(2)}(x) =  g_2(x) G(x) = \frac{g_2(x)}{g_1(x)}H(x) 
\label{recursiveencoding}
\end{eqnarray}
Two different recursive codes can be defined by permuting the two 
polynomials $g_1$ and $g_2$.\\
It is easy to show that Eq. (\ref{recursiveencoding}) is equivalent to 
\begin{eqnarray}
\sigma_i(\tg)  = \Sigma_0(i)  
= \tau_i\prod_{j=1}^r \Sigma_j(i)^{\kappa(j;1)}  &=& 
\tau_i\prod_{j=1}^r (\sigma_{i-j})^{\kappa(j;1)}
\label{recursivefeedback}
\end{eqnarray}
From Eq.(\ref{recursivefeedback}) it follows that:
\begin{eqnarray}
\tau_i(\sg) = \prod_{j=0}^r (\sigma_{i-j})^{\kappa(j;1)} & = &
J^{(1)}_i
\label{recursiveregister}
\end{eqnarray}
Because of the last equality in Eq.(\ref{recursiveregister}) a part of
the encoded message (in the recursive case) is always the message
itself.\\
We shall now consider decoding. 
Using the method explained in the Introduction the probability
distribution  of the register sequence conditional to some ouput can 
be written as the Boltzmann weight of a spin model with random
couplings. The Hamiltonian of this model is:
\begin{eqnarray}
H(\sg;\Jg^{\mbox{out}}) & = & 
-\sum_{i=1}^N \left\{ B(J^{(1),\mbox{out}}_i)\prod_{j=0}^r
(\sigma_{i-j})^{\kappa(j;1)}+B(J^{(2),\mbox{out}}_i)\prod_{j=0}^r
(\sigma_{i-j})^{\kappa(j;2)}\right\}\nonumber\\
\end{eqnarray}
where $B(\cdot   )$ is defined in Eq.(\ref{DefinizioneB}).\\
For convolutional codes the model is one dimensional with two types of
couplings. The range of the interaction coincide with the range of the
code. 
The alert reader will notice that the Hamiltonian is expressed as a function 
of the spins of the register sequence $\sigma_i $, 
instead of the source sequence $\tau_i $ used 
in the introduction. For non recursive codes $\sigma_i = \tau_i $. 
For recursive codes $\tau_i $ is given by Eq.(\ref{recursiveregister}), 
i.e. in this last case decoding can be thought of as the computation of 
an expectation value of a composite operator. However the spin Hamiltonian is 
the same for both the recursive and not recursive codes.\\
We define the decoding at arbitrary temperature $T\equiv 1/\beta$ as follows:
\begin{eqnarray}
\tau_i ^{\beta} &\equiv& \mbox{sign}(\<\tau_i(\sg)\>_{\beta})
\label{decbeta}\\
\<O(\sg)\>_{\beta}&\equiv&\frac{1}{Z(\Jg^{\mbox{out}};\beta)}
\sum_{\sg}O(\sg)\exp\{-\beta H(\sg;\Jg^{\mbox{out}})\}\label{decbetaexp}
\end{eqnarray}
where the expression for $\tau_i(\sg)$ is given by
Eq.(\ref{recursiveregister}) or by Eq.(\ref{RegistroNonRicorsivo})
depending whether the code is recursive or not.\\
As seen in the introduction there are two widely used decoding strategies:
\begin{itemize}
\item Maximum Likelihood decoding which consists 
in finding the most probable sequence of bits and
corresponds to the choice $\beta = \infty$ in Eq.(\ref{decbeta}): $\tau ^{ML}_i
\equiv \tau ^{\beta=\infty}_i$.
\item  Maximum A posteriori Probability decoding which consists 
in finding the most probable sequence of bits and
corresponds to the choice $\beta = 1$ in Eq.(\ref{decbeta}): $\tau ^{MAP}_i
\equiv \tau ^{\beta=1}_i$. This is the strategy which enters in turbo 
decoding. 
\end{itemize}
Both this strategies can be implemented in a very efficient way using
the transfer matrix technique. The corresponding algorithms are known
in communication theory as the Viterbi algorithm \cite{AlgoritmoViterbi}
for the $\beta = \infty$
case and the BCJR algorithm \cite{AlgoritmoBCJR} 
for the $\beta=1$ case. The complexity of
these algorithms grows like $N 2^r$.\\
The use of the register sequence (i.e. of the $\sigma$ variables) 
makes evident the similarity between recursive and nonrecursive
codes: they correspond to the same spin model. This implies e.g.
that, if zero temperature decoding is adopted, the probability
of transmitting a message without errors is the same with the two codes.\\
In the limit $r\to\infty$ it is possible to construct convolutional
codes corresponding to spin models
with infinite connectivity and couplings between an infinite number of
spins. They should allow to transmit without errors when the noise is
low enough. In practice, because of the growing complexity of the
transfer matrix algorithm, a compromise between low $r$'s
(which are simpler to decode) and high $r$'s (which show better
performances) must be found. The values of $r$ used in practical cases
are between $2$ and $7$.\\ 
We can write the decoding strategy in terms of the message (i.e. of
the $\tau$ variables) without making use of the register
sequence (i.e. of the $\sigma$ variables):
\begin{eqnarray}
\tau_i ^{\beta} = \mbox{sign}\left\{\frac{1}{Z(\Jg^{\mbox{out}};\beta)}
\sum_{\tg}\tau_i\exp\{-\beta H(\sg(\tg);\Jg^{\mbox{out}})\} \right\}
\end{eqnarray}
For non recursive codes, because of Eq.(\ref{RegistroNonRicorsivo}), 
 things remain unchanged. 
However, for recursive codes, since Eq.(\ref{recursiveregister}) cannot be 
inverted in a local way, we obtain a non local 
Hamiltonian.\\
As a simple illustration of this observation we can consider the Hamiltonian
corresponding to the code \ref{CodiceSemplice}:
\begin{eqnarray}
H^{(\mbox{a})}(\sg;\Jg) & = & 
-\sum_{i=1}^NB(J^{(1),\mbox{out}}_i)\sigma_i\sigma_{i-1}-
\sum_{i=1}^NB(J^{(2),\mbox{out}}_i)\sigma_i\\
H^{(\mbox{a})}(\sg(\tg);\Jg) & = & 
-\sum_{i=1}^NB(J^{(1),\mbox{out}}_i)\tau_i-
\sum_{i=1}^NB(J^{(2),\mbox{out}}_i)\prod_{j=1}^i\tau_j
\end{eqnarray}
For less simple codes we define the 
numbers $\rho(j)\in\{0,1\}$ as follows:
\begin{eqnarray}
\frac{g_2(x)}{g_1(x)} \equiv \sum_{j=0}^{\infty}\rho(j)x^j\pmod{2}
\end{eqnarray}
we get
\begin{eqnarray}
H(\sg(\tg);\Jg) & = & 
-\sum_{i=1}^NB(J^{(1),\mbox{out}}_i)\tau_i-
\sum_{i=1}^NB(J^{(2),\mbox{out}}_i)\prod_{j=0}^{i-1}(\tau_{i-j})^{\rho(j)}
\end{eqnarray}
Written in this form recursive codes look very different from non
recursive ones with the same range.  
If $g_2(x)$ is not divisible by $g_1(x)$ the corresponding spin
models have infinite connectivity and interactions with infinite
range; they are similar, in this respect, 
to $r=\infty$ non recursive codes.\\
Neverthless they do not behave, in general, radically better
than the non recursive codes with the same range because there exists, 
as we have shown, a change of variables (from $\tau$ to
$\sigma$) which makes the model local. 

%  
%****************************************************************************
%
\section{Turbo codes.}

A turbo code is defined by the choice of a convolutional code
and of a permutation of $N$ objects. We use for the permutation 
the following notation:
\begin{eqnarray}
P: \{1,\dots,N\} &\rightarrow &\{1,\dots,N\}\\
i&\mapsto & P(i)
\end{eqnarray}
and we shall denote by $P^{-1}$ the inverse permutation
($P(P^{-1}(i))=P^{-1}(P(i))=i$).\\
The basic idea is to apply the 
permution $P$ to the source sequence $\tg$ to produce a new sequence 
$\tg ^P $. Obviously $\tg ^P $ does not carry any new information
because $P$ is known. 
Both sequences $\tg$ and $\tg ^P $ are the inputs to two set of 
registers, each one implementing a convolutional encoding. 
In this way the rate of the code is decreased 
(i.e. greater redundancy). One can increase the rate by erasing some 
of the outputs \cite{PrimoBerrou}, 
but we will not consider this possibility in this paper.\\
The properties of the system can strongly depend on
the choice of the permutation. Permutations ``near'' the identity give very
bad codes. We shall think to a ``good'' permutation as to a random
permutation. In the limit $N\to\infty$ they are ``far'' from the identity
 with probability one. We shall discuss this point later in this section.\\ 
We illustrate this idea with the example of a rate $1/2$ recursive 
convolutional code, defined by the constants
$\kappa(j;1)$ and $\kappa(j;2)$. The two register sequences are: 
\begin{eqnarray}
\sigma_i ^{(1)}\equiv\sigma_i(\tg)&\Rightarrow& 
\tau_i = 
\prod_{j=1}^r (\sigma ^{(1)}_{i-j})^{\kappa(j;1)}\equiv 
\epsilon_i(\sg^{(1)})\label{TurboRegistro1}\\
\sigma_i ^{(2)}\equiv\sigma_i(\tg^P)&\Rightarrow&
\tau ^P_i = 
\prod_{j=1}^r (\sigma ^{(2)}_{i-j})^{\kappa(j;1)}\equiv
\epsilon_i(\sg^{(2)})\label{TurboRegistro2}
\end{eqnarray}
where $\tg^P$ is the permuted message ($\tau
^P_i\equiv\tau_{P(i)}$).\\
The relation between the two register sequences is rather involved and
nonlocal for a general choice of the permutation. Moreover 
$\sigma ^{(1)}_i$ can be expressed only in terms of a large number of 
$\sigma ^{(2)}_j$'s.
The identity permutation is clearly an exception since in this case 
$\sigma ^{(1)}_i = \sigma ^{(2)}_i$.\\ 
Let us consider as an example the code \ref{CodiceSemplice}:
\begin{eqnarray}
\sigma ^{(1)}_i = \prod_{j=1}^i\sigma ^{(2)}_{P(j)}\sigma ^{(2)}_{P(j)-1}
\label{DueRegistri}
\end{eqnarray}
It is simple to show that, for a random permutation, the number of
different $\sigma ^{(2)}$'s in the product on the r.h.s. of 
Eq.(\ref{DueRegistri}) is of order $O(N)$.\\
The turbo code defined by the permutation $P$ and by the numbers
$\kappa(j;1)$  and $\kappa(j;2)$
has rate $1/3$ and the encoded message has the following form:
\begin{eqnarray}
J^{(0)}_i & \equiv & \prod_{j=0}^r
(\sigma_{i-j}(\tg))^{\kappa(j;1)}=\tau_i
\label{DefinizioneTurbo1}\\
J^{(1)}_i & \equiv & \prod_{j=0}^r (\sigma_{i-j}(\tg))^{\kappa(j;2)}
\label{DefinizioneTurbo2}\\
J^{(2)}_i & \equiv & \prod_{j=0}^r (\sigma_{i-j}(\tg^P))^{\kappa(j;2)}
\label{DefinizioneTurbo3}
\end{eqnarray}
It turns out that it is convenient to write the corresponding Hamiltonian 
as a function of both register sequences. This introduces new degrees of 
freedom and the Hamiltonian is a function of $2N$ instead of $N$
spin. The unwanted degrees of freedom are eliminated by imposing 
 the constraint $\tau^{P}_i = \tau_{P(i)} $. This constraint can be 
written in terms of the $\sigma$'s using Eqs.(\ref{TurboRegistro1}) 
and (\ref{TurboRegistro2}).
The probability distribution for the register sequences can then be written
as:
\begin{eqnarray}
P(\sg^{(1)},\sg^{(2)}|\Jg^{\mbox{out}}) & = &
\frac{1}{Z(\Jg^{\mbox{out}})}
\prod_{i=1}^N\delta(\epsilon_{P(i)}(\sg^{(1)}),
\epsilon_{i}(\sg^{(2)}))\nonumber\\
&&\exp\{-H(\sg^{(1)},\sg^{(2)};\Jg^{\mbox{out}})\}\label{probabilitaturbo}\\
 H(\sg^{(1)},\sg^{(2)};\Jg^{\mbox{out}})& = & 
-\sum_{i=1}^N \left\{ B(J^{(0),\mbox{out}}_i)\prod_{j=0}^r
(\sigma ^{(1)}_{i-j})^{\kappa(j;1)}+\right.\nonumber\\
&&\phantom{-\sum_{i=1}^N \left\{ \right.}
 +B(J^{(1),\mbox{out}}_i)\prod_{j=0}^r(\sigma
^{(1)}_{i-j})^{\kappa(j;2)}+\label{probabilitaturbo2}
\label{TurboHamiltonian}\\
&&\phantom{-\sum_{i=1}^N \left\{ \right.}\left.+B(J^{(2),\mbox{out}}_i)
\prod_{j=0}^r(\sigma ^{(2)}_{i-j})^{\kappa(j;2)}\right\}
\nonumber
\end{eqnarray}
where $\delta(i,j)$ is the ordinary Kronecker function. 
In this way the probability distribution is a local
function of the spin variables $\sigma^{(1)}$ and $\sigma ^{(2)}$.\\ 
We shall call the code defined by 
Eqs.(\ref{DefinizioneTurbo1},\ref{DefinizioneTurbo2},\ref{DefinizioneTurbo3}) 
a non recursive turbo code if
$\kappa(j;1)=\delta_{j,0}$ and a recursive turbo code otherwise.
Recursive turbo codes are the ones usually called turbo codes in
communication theory.\\
The probability distribution for the recursive turbo code  
(\ref{probabilitaturbo}) can't be written in terms of one of the two
register sequences $\sg ^{(1)}$ or $\sg ^{(2)}$ without producing
large connectivities (see Eq.(\ref{DueRegistri})).\\
If $P$ is the identity permutation then $\sg^{(1)} = \sg^{(2)}$ and
the code becomes a convolutional one with the same rate ($1/3$) and
the same generating polynomials. We shall use the convolutional code
obtained in this way as a standard comparison term for the
performances of turbo codes (see Figs.(\ref{trecneargraf}-\ref{trecgraf})). 
The outcome of this
comparison (i.e. recursive turbo codes have a much lower error probability than
convolutional codes) demonstrates the importance of the choice of the 
permutation.\\
For non recursive turbo codes the two register sequences are 
related simply by a permutation:
\begin{eqnarray}
\sigma ^{(1)}_i = \sigma_i(\tg) = \sigma_{P^{-1}(i)}(\tg^P) =  
\sigma ^{(2)}_{P^{-1}(i)}
\end{eqnarray}
and
\begin{eqnarray}
P_{\mbox{\small{non-rec}}}(\sg^{(1)}|\Jg^{\mbox{out}}) & = &
\frac{1}{Z(\Jg^{\mbox{out}})}
\exp\{-H(\sg^{(1)},(\sg^{(1)})^P;\Jg^{\mbox{out}})\}\\
H(\sg^{(1)},\sg^{(2)};\Jg^{\mbox{out}}) &\equiv &-\sum_{i=1}^N\left\{
B(J^{(0)\mbox{out}}_i)\sigma_i ^{(1)}+
B(J^{(1)\mbox{out}}_i)\prod_{j=0}^r
(\sigma_{i-j}^{(1)})^{\kappa(j;2)}+\right.\nonumber\\
&&\phantom{-\sum_{i=1}^N}\left.B(J^{(2)\mbox{out}}_i)\prod_{j=0}^r
(\sigma_{i-j}^{(2)})^{\kappa(j;2)}\right\}
\end{eqnarray}
so that the spin model corresponding to this type of code has a finite
connectivity $c = 1+2\sum_{j=0}^r\kappa(j;2)$.\\
This finite versus infinite connectivity is the essential difference
between non recursive and recursive turbo codes and explains why 
recursive turbo codes are so better and why they can achieve
zero error probability for low enough noise.\\
We now discuss decoding. There is no exact decoding algorithm for 
turbo codes. Berrou et al. have proposed a very ingenious algorithm, 
called turbo decoding, which is thought to be approximate. 
Turbo decoding is an iterative procedure. 
At each step of the iteration, one considers one of the two 
chains, i.e. either the couplings $\Jg^{(0)}$ and  $\Jg^{(1)}$ 
or  $\Jg^{(0)}$ and  $\Jg^{(2)}$ and proceeds to MAP decoding. 
The information so obtained  is injected to the next step by adding  
appropriate external  fields to the Hamiltonian. The algorithm
terminates if a fixed point is reached.\\
In order to explain the algorithm more precisely, we introduce the following 
expectation values:
\begin{eqnarray}
\Xi_i[\Bg,\Bg' ] &\equiv &\frac{1}{Z}
\sum_{\sg}\epsilon_i(\sg)\exp\left\{\sum_{i=1}^N
B_i\prod_{j=0}^r(\sigma_{i-j})^{\kappa(j;1)}+
\sum_{i=1}^N B'_i\prod_{j=0}^r(\sigma_{i-j})^{\kappa(j;2)}
\right\}
\end{eqnarray}
The $\Xi_i$'s can be computed in an efficient way by using
the finite temperature transfer matrix algorithm.
They are the expectation values of the operator defined by 
Eqs.(\ref{TurboRegistro1}) or (\ref{TurboRegistro2}).\\
Then we introduce the iteration variables:
\begin{eqnarray}
\thg^{(m)}(t)\equiv (\theta ^{(m)}_1(t),\dots,\theta ^{(m)}_N(t))\\
\Gg^{(m)}(t)\equiv (\Gamma ^{(m)}_1(t),\dots,\Gamma ^{(m)}_N(t))
\end{eqnarray}
for $m=1,2$.\\
In terms of these variables the iteration reads
\begin{eqnarray}
\theta ^{(1)}_i(t+1)  &=& 
\Xi_i[\Bg^{(0)}+\Gg^{(1)}(t),\Bg^{(1)}] 
\label{turboiter1}\\
\theta ^{(2)}_i(t+1)  &=& 
\Xi_i[\Bg^{(0),P}+\Gg^{(2)}(t),\Bg^{(2)}] 
\label{turboiter2}\\
\Gamma ^{(1)}_i(t+1) & = & \arctanh\left[\theta
^{(2)}_{P^{-1}(i)}(t+1)\right]-\Gamma^{(2)}_{P^{-1}(i)}(t)
-B^{(0)}_i\label{turboiter3}\\
\Gamma ^{(2)}_i(t+1) & = & \arctanh\left[\theta
^{(1)}_{P(i)}(t+1)\right]-\Gamma^{(1)}_{P(i)}(t)
-B^{(0)}_{P(i)}
\label{turboiter4}
\end{eqnarray}
with
\begin{eqnarray}
\Bg^{(m)}\equiv (B(J_1 ^{(m)\mbox{out}}),\dots,B(J_N ^{(m)\mbox{out}}))
 \ ; \ m=0,1,2
\end{eqnarray}
and $B^{(0),P}_i \equiv B^{(0)}_{P(i)}$.\\
The meaning of the previous equations is the following.
 The $\theta_i$ are expectation values of 
a sequence of operators 
which can take only values $\pm   1$, computed independently for every 
element of the sequence. The information contained in $\theta_i$ can therefore 
be represented by an ``external field'' 
$\Gamma_i$ such that $\theta_i = \tanh \Gamma_i$. In order to avoid double 
counting of information one substracts the external fields of the previous 
iteration as shown in  Eqs.(\ref{turboiter3},\ref{turboiter4}).\\
Hopefully the iteration converges to a fixed point:
\begin{eqnarray}
\lim_{t\to\infty} \theta ^{(1)}_i(t) = 
\lim_{t\to\infty} \theta ^{(2)}_{P^{-1}(i)}(t) \equiv \theta ^{*}_i
\end{eqnarray}
The decoded message is obtained as follows:
\begin{eqnarray}
\tau ^{TURBO}_i \equiv \mbox{sign}(\theta ^{*}_i)
\label{turbodecodingdefinition}
\end{eqnarray}
The system described by 
Eq.(\ref{probabilitaturbo},\ref{probabilitaturbo2}) is
seen in turbo decoding as the union of two one dimensional
subsystem. Each subsystem acts on the other one through a magnetic
field (in the non recursive case) or through an additional coupling
(in the recursive case).\\
To get some insight of
Eqs.(\ref{turboiter1}-\ref{turboiter4}) we define the
free energy functionals $F^{(1)}$ and $F^{(2)}$:
\begin{eqnarray}
{\cal Z}^{(1)}[\Gg] &=& \sum_{\sg}\exp\left\{
\sum_{i=1}^N
(B(J^{(0)}_i)+\Gamma_i)\prod_{j=0}^r(\sigma_{i-j})^{\kappa(j;1)}+
\sum_{i=1}^N
B(J^{(1)}_i)\prod_{j=0}^r(\sigma_{i-j})^{\kappa(j;2)}\right\}
\nonumber\\
\\
{\cal Z}^{(2)}[\Gg] &=& \sum_{\sg}\exp\left\{
\sum_{i=1}^N
(B(J^{(0)}_{P(i)})+\Gamma_{P(i)})\prod_{j=0}^r(\sigma_{i-j})^{\kappa(j;1)}+
\sum_{i=1}^N
B(J^{(2)}_i)\prod_{j=0}^r(\sigma_{i-j})^{\kappa(j;2)}\right\}
\nonumber\\
\\
F^{(m)}[\thg] & \equiv & \left.\sum_{i=1}^N\theta_i\Gamma_i-
\log\left({\cal Z}^{(m)}[\Gg]\right)
\right|_{\theta_i = \frac{\partial\log({\cal Z}^{(m)})}{\partial\Gamma_i} }
\end{eqnarray}
It is then simple to show that $\theta ^{*}$ is a solution of the
equation:
\begin{eqnarray}
\frac{\partial}{\partial \theta_i}{\cal
F}^{\mbox{\footnotesize{turbo}}}[\thg] & = & 0
\end{eqnarray}
where
\begin{eqnarray}
{\cal F}^{\mbox{\footnotesize{turbo}}}[\thg] &\equiv&
F^{(1)}[\thg]+F^{(2)}[\thg]-F_0[\thg]\label{turbofreeenergy}\\
F_0[\thg]&\equiv&\sum_{i=1}^N\left\{-B^{(0)}_i\theta_i-s(\theta_i)\right\}\\
s(x)&\equiv &-\left(\frac{1+x}{2}\right)\log\left(\frac{1+x}{2}\right)-
\left(\frac{1-x}{2}\right)\log\left(\frac{1-x}{2}\right)
\end{eqnarray}
Eq.(\ref{turbofreeenergy}) is an approximation to 
the true free energy functional of the total system which is given by:
\begin{eqnarray}
{\cal Z}[\Gg] &\equiv &\sum_{\sg^{(1)}}\sum_{\sg^{(2)}}
\prod_{i=1}^N\delta(\epsilon_{P(i)}(\sg^{(1)}),
\epsilon_{i}(\sg^{(2)}))\nonumber\\
&&\phantom{\sum_{\sg^{(1)}}\sum_{\sg^{(2)}}}
\exp\left\{-H(\sg^{(1)},\sg^{(2)};\Jg^{\mbox{out}})+
\sum_{i=1}^N \Gamma_i\epsilon_i(\sg^{(1)})\right\}\\
&&{\cal F}[\thg;\Jg^{(0)},\Jg^{(1)},\Jg^{(2)}]
\equiv  \left.\sum_{i=1}^N\theta_i\Gamma_i-
\log\left({\cal Z}[\Gg]\right)
\right|_{\theta_i = \frac{\partial\log({\cal Z})}{\partial\Gamma_i} }
\end{eqnarray}
where $H(\sg^{(1)},\sg^{(2)};\Jg^{\mbox{out}})$ is given in
Eq.(\ref{probabilitaturbo2}).\\ 
It is then evident that
\begin{eqnarray}
{\cal F}^{\mbox{\footnotesize{turbo}}}[\thg] = 
{\cal F}[\thg;\Jg^{(0)},\Jg^{(1)},\zb]
+{\cal F}[\thg;\Jg^{(0)},\zb,\Jg^{(2)}]-
{\cal F}[\thg;\Jg^{(0)},\zb,\zb]
\end{eqnarray}
i.e. turbo decoding neglects terms of order 
$B(J^{(1)}_{i_1})B(J^{(2)}_{i_2})$.\\
\section{Replica approach.}
\label{ReplicaSection}
We would like to compute the error probability per bit. 
As explained in the introduction, in the case of a symmetric transmission
channel, it is enough to compute the magnetization in the case of all inputs 
$\tau_i = 1$. The error probability per bit is given by the  
probability of a local magnetization being negative.\\
The similarity of the Hamiltonian (\ref{TurboHamiltonian}) 
with the Hamiltonians of disordered spin systems is obvious. 
The disorder in the case of turbo codes has two origins. 
One is due to the (random) permutation  
which defines the particular code. The other is more conventional 
and is related to the randomness of the couplings which is due to the 
transmission noise. As usual in disordered systems, we can only 
compute the average over disorder and for that we have to introduce replicas.\\
Let us define the expectation value of the operator
$\tau_i(\sg)$ defined in
Eqs.(\ref{TurboRegistro1},\ref{TurboRegistro2})
with respect to the probability distribution given by 
Eqs.(\ref{probabilitaturbo},\ref{probabilitaturbo2}):
\begin{eqnarray}
\Theta_i[\Jg^{\mbox{out}},P]\equiv \sum_{\sg^{(1)}}\sum_{\sg^{(2)}}
\epsilon_i(\sg^{(1)})P(\sg^{(1)},\sg^{(2)}|\Jg^{\mbox{out}})
\end{eqnarray}
The statistical properties of a turbo code can be derived from the
probability distribution of this expectation value:
\begin{eqnarray}
{\cal P}_i(\theta|P)\equiv
\int \!dQ[\Jg^{\mbox{out}}]\, 
\delta\left(\theta-\Theta_i[\Jg^{\mbox{out}},P]\right)
\quad
i=1,\dots,N
\label{DistribuzioneSito}
\end{eqnarray}
where
\begin{eqnarray}
dQ[\Jg^{\mbox{out}}] = \prod_{n=0}^2\prod_{i=1}^N Q(J^{(n)}_i|+1)dJ^{(n)}_i
\end{eqnarray}
Then we define the average distribution 
\begin{eqnarray}
\overline{{\cal P}}(\theta)\equiv
\frac{1}{N!}\sum_{P}    
\int \!dQ[\Jg^{\mbox{out}}]\;
\delta\left(\theta-\Theta_i[\Jg^{\mbox{out}},P]\right)
\end{eqnarray}
where the sum runs over all possible permutations.
$\overline{{\cal P}}(\theta)$
 is expected not to depend upon the site $i$ in the thermodynamic
limit ($N\to\infty$).\\
The average error probability per bit is given by
\begin{eqnarray}
\overline{P_e}\equiv\int_{-\infty}^0 \!d\theta\;\overline{{\cal P}}(\theta)
\end{eqnarray}
In any case $\overline{P_e}$ is an upper bound for the error
probability of the ``best'' code (i.e. the one buildt with the
permutation which yields the lowest error probability).\\
The replicated partition function is given by:
\begin{eqnarray}
\overline{Z^n}&\equiv&\frac{1}{N!}\sum_{P}
\int \!dQ[\Jg^{\mbox{out}}]\;\sum_{\{\sg^{(1),a}\}}\sum_{\{\sg^{(2),a}\}}
\prod_{a=1}^n
\prod_{i=1}^N
\delta(\epsilon_{P(i)}(\sg^{(1),a}),\epsilon_{i}(\sg^{(2),a}))\nonumber\\
&&\phantom{\sum_{P}w(P)\int \!dQ[\Jg^{\mbox{out}}]\;}
\exp\left\{-\sum_{a=1}^n H(\sg^{(1),a},\sg^{(2),a};\Jg^{\mbox{out}})
\right\}
\end{eqnarray}
The average over permutations can be done by introducing a matrix
representation of the permutation
\begin{eqnarray}
C^P_{ij}\equiv\delta_{i,P(j)} \quad  ; \quad i,j = 1,\dots,N
\end{eqnarray}
To sum over permutations, one sums over all matrices $C^P_{ij} = 0 $ or $ 1 $ 
 with the constrain $\sum_i C^P_{ij} = 
\sum_i C^P_{ij} = 1 $.
One may use the identity $ \delta ( \sigma, \tau ) = ( 1 + \sigma \tau )/2 $ 
to write 
\begin{eqnarray}
\prod_{a=1}^n\prod_{i=1}^N\delta(\epsilon_{P(i)}(\sg^{(1),a}),
\epsilon_{i}(\sg^{(2),a})) = \prod_{a=1}^n\prod_{i,j=1}^N
\left[\left(1-\frac{1}{2}C^P_{ij}\right)
+\frac{1}{2}C^P_{ij}\epsilon_i(\sg^{(1),a})\epsilon_j(\sg^{(2),a})\right]  
\end{eqnarray} 
It can be shown \cite{Mio} that the ``effective action'' 
 which is obtained in this way describes two
one-dimensional models coupled by a mean field-like interaction.\\
This is easily seen by making use of the occupation densities 
\cite{OccupationDensity} defined below:
\begin{eqnarray}
c_m(\underline{\epsilon})&\equiv &\frac{1}{N}\sum_{i=1}^N 
\delta_{\underline{\epsilon},\underline{\epsilon}^{(m)}_i}\\
\underline{\epsilon} &\equiv&
(\epsilon^1,\dots,\epsilon^a,\dots,\epsilon^n)\in\{-1,+1\}^n\nonumber\\
\underline{\epsilon}^{(m)}_i&\equiv &
(\epsilon_i(\sg^{(m),1}),\dots,
\epsilon_i(\sg^{(m),a}),\dots,
\epsilon_i(\sg^{(m),n}))\nonumber
\end{eqnarray}
The resulting replicated partition function reads \cite{Mio}:
\begin{eqnarray}
\overline{Z^n}&\equiv&
\int \!dQ[\Jg^{\mbox{out}}]\;\sum_{\{\sg^{(1),a}\}}\sum_{\{\sg^{(2),a}\}}
\prod_{\underline{\epsilon}}
\delta_{Nc_1(\underline{\epsilon}),Nc_2(\underline{\epsilon})}\\
&&\phantom{\int \!dQ[\Jg^{\mbox{out}}]\;}
\exp\left\{-\sum_{a=1}^n H(\sg^{(1),a},\sg^{(2),a};\Jg^{\mbox{out}})+
N\sum_{\underline{\epsilon}}c_1(\underline{\epsilon})
\log c_1(\underline{\epsilon})\right\}\nonumber
\end{eqnarray}
We briefly report here the main results of this approach 
for the gaussian channel described by Eq.(\ref{GaussianChannel}). 
A detailed analysis will be presented elsewhere \cite{Mio}.\\
For recursive turbo codes there exists a low noise phase $w^2<w_c^2$
where the error probability vanishes in the thermodynamic limit
(i.e. for infinitely long sequences). In this phase the model is
completely ordered:
\begin{eqnarray}
\overline{{\cal P}}(\theta) = \delta(\theta-1)
\end{eqnarray}
A local stability analysis yields the critical value $w^2_{loc}$ such
that for $w^2>w^2_{loc}$ the no-error phase is destroyed by small
fluctuations. Clearly $w^2_{loc}\ge w^2_{c}$. We computed $w^2_{loc}$ 
for the two cases listed below.\\
 For both the rate is $R=1/3$ so that the Shannon 
noise threshold as given by 
Eq.(\ref{GaussianCapacity}) is
$w^2_{Shannon} = 1/(2^{2/3}-1)\simeq -2.31065 \mbox{ db}$. Error free
communication can take place only for $w^2 <w^2_{Shannon}$.
\begin{itemize}
\item For model \ref{CodiceSemplice}, defined by 
Eqs.(\ref{CodiceSemplice1}),(\ref{CodiceSemplice2}) one gets $w_{loc}^2 =
1/\ln 4\simeq 1.41855 \mbox{ db}$. 
\item For model \ref{CodiceTipico}, defined by 
Eqs.(\ref{CodiceTipico1}),(\ref{CodiceTipico2}) one obtains $w_{loc}^2 =
-1/(2\ln x_c)$ where $x_c \simeq 0.741912\dots$ is the only real
solution of the equation
\begin{eqnarray}
2x^5+x^2 = 1
\end{eqnarray}
The resulting value $w^2_{loc}\simeq -2.23990 \mbox{ db}$ is quite near to
the Shannon threshold. 
\end{itemize}
%
%%%%%%%%%%%%%%%%%%%%%%%%%%%%%%%%%%%%%%%%%%%%%%%%%%%%%%%%%%%%%%%%%
%
\section{Discussion.}
\begin{figure}
\centerline{
\epsfig{figure=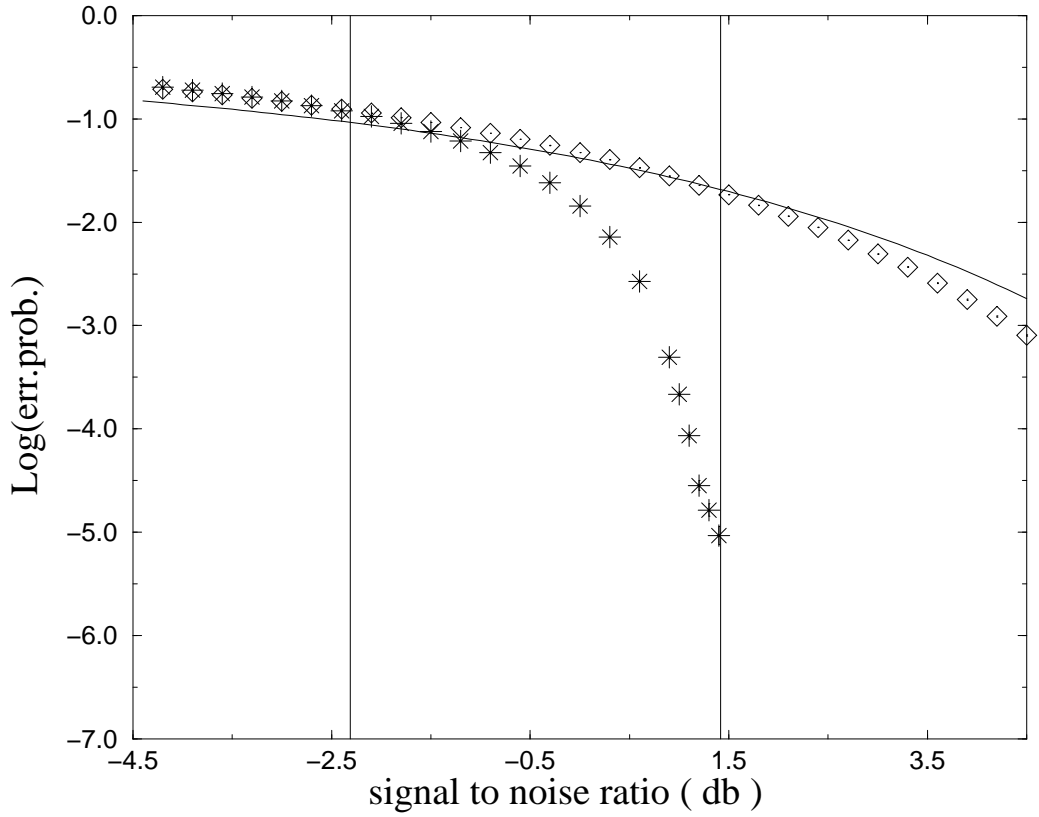,angle=-90,
width=1.0\linewidth}
}
\caption{Numerical results for the error probability per bit of the
recursive turbo code buildt from the convolutional code \ref{CodiceSemplice}
(cfr. Eqs.(\ref{CodiceSemplice1}),(\ref{CodiceSemplice2})).
Stars ($\ast$) refer to the turbo code, diamonds ($\Diamond$) 
to the convolutional code obtained by setting the permutation 
$P$ equal to the identity permutation, and the continuous line to the
uncoded message.
The leftmost vertical line is located at the Shannon
thresold ($w^2 = w^2_{Shannon}$) while the rightmost at the thresold
of local stability ($w^2 = w^2_{loc}$, see
Sec.(\ref{ReplicaSection})).}
\label{trecneargraf}
\end{figure}

\begin{figure}
\centerline{
\epsfig{figure=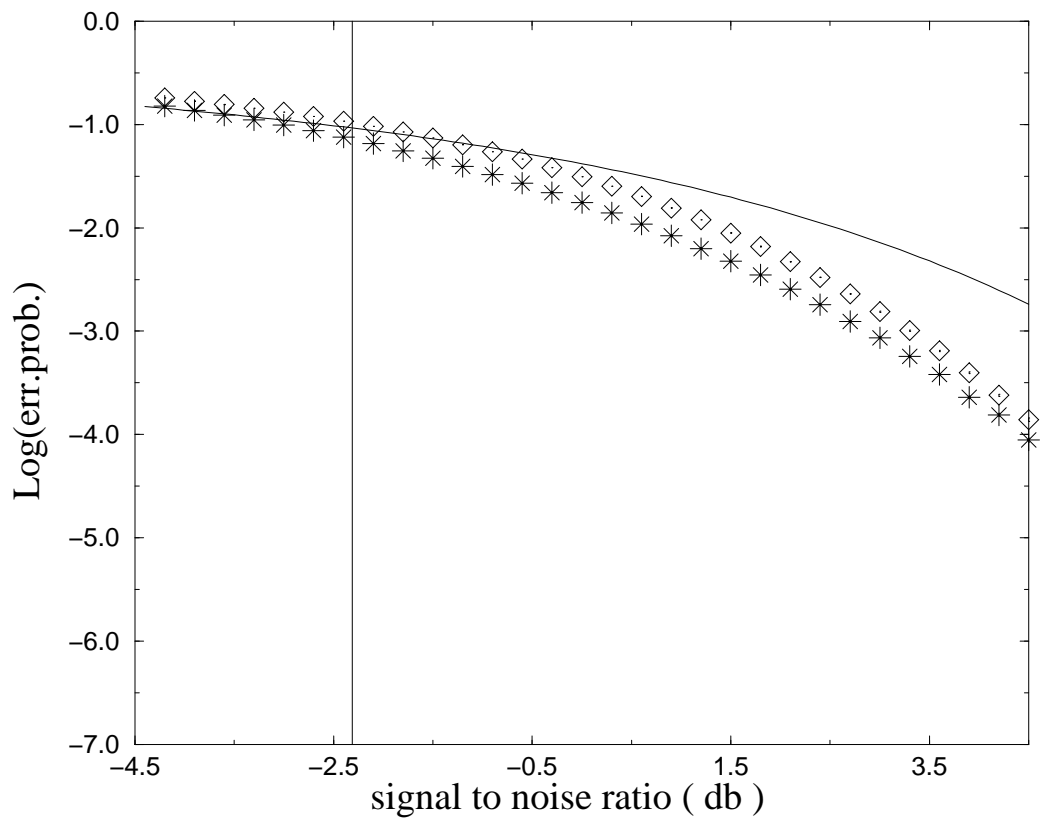,angle=-90,
width=1.0\linewidth}
}
\caption{Numerical results for the error probability per bit of the
non recursive turbo code buildt from the convolutional code 
\ref{CodiceSemplice}
(cfr. Eqs.(\ref{CodiceSemplice1}),(\ref{CodiceSemplice2})).
The symbols have the same meaning explained in the caption of 
Fig.(\ref{trecneargraf}).}
\label{turboneargraf}
\end{figure}

\begin{figure}
\centerline{
\epsfig{figure=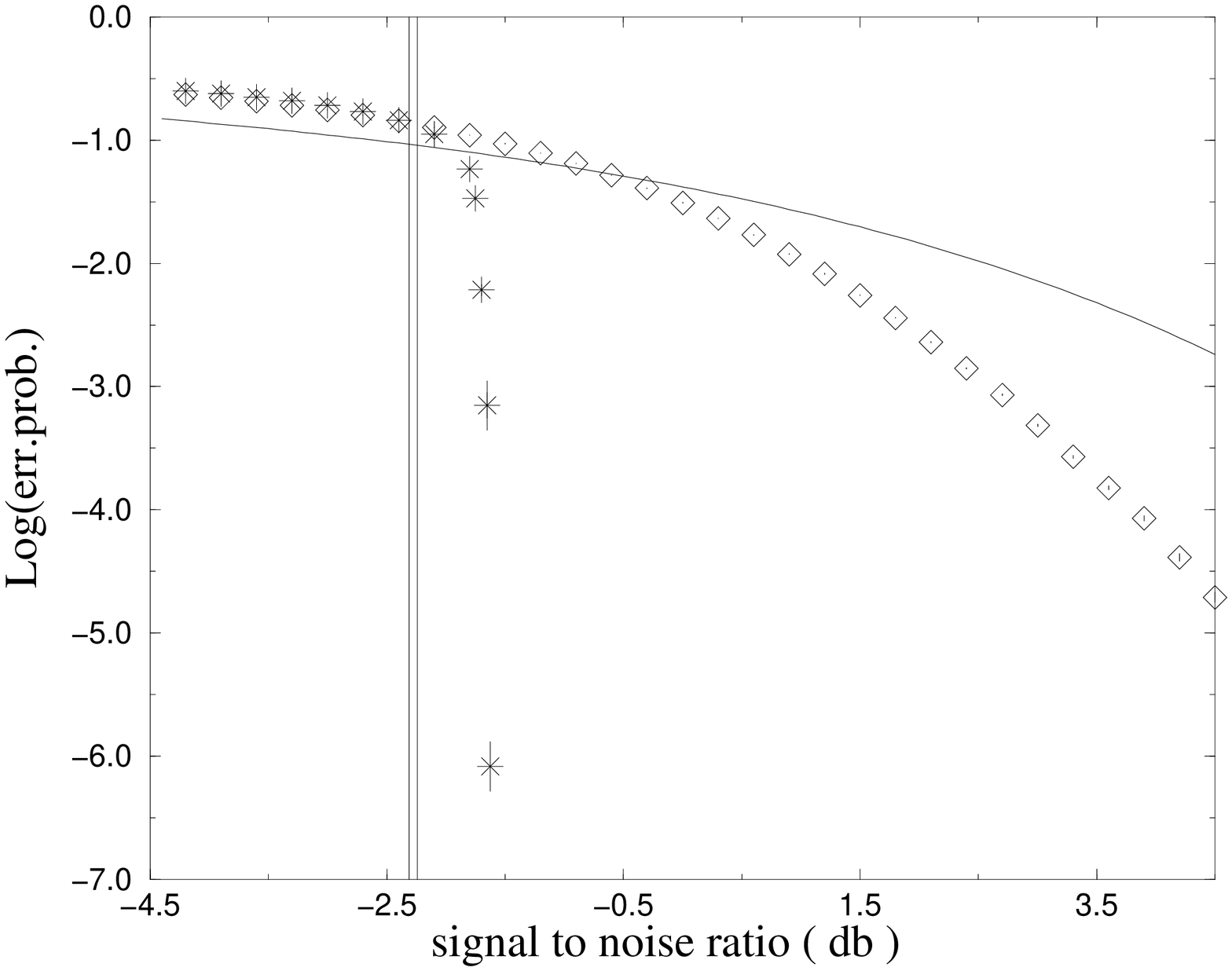,angle=-90,
width=1.0\linewidth}
}
\caption{Numerical results for the error probability per bit of the
recursive turbo code buildt from the convolutional 
code \ref{CodiceTipico} 
(cfr. Eqs.(\ref{CodiceTipico1}),(\ref{CodiceTipico2})).
For the meaning of various symbols see the caption of 
Fig.(\ref{trecneargraf}). Notice that in this case 
$w^2_{loc}\sim w^2_{Shannon}$.}
\label{trecgraf}
\end{figure}

We formulated turbo codes as a spin model Hamiltonian and we 
obtained new results using the replica method. 
It is well known that this method is not mathematically rigorous.  
So it is natural to question the validity of  
our results. For this purpose we have carried out numerical simulations 
of the following codes: 
the recursive turbo code corresponding to the convolutional 
code \ref{CodiceSemplice} of Sec.(\ref{ConvolutionalSection}), its
error probability is reported in Fig.(\ref{trecneargraf});
the non recursive turbo code obtained by permuting the generating
polynomials of the previous one (see Fig.(\ref{turboneargraf})); 
the recursive turbo code corresponding to the code
\ref{CodiceTipico} of the same section (see Fig.(\ref{trecgraf})). 
We used the Berrou et al. turbo decoding 
algorithm and averaged over $200$ to $500$ realizations of the disorder.\\
The first conclusion is that recursive turbo codes are much 
better codes than non recursive ones. Furthemore our results 
for recursive turbo codes are 
compatible with the existence of a threshold $w_c^2$ such that 
for $ w^2 < w_c^2 $ the error probability per bit is zero, while no such 
threshold seems to exist for non recursive codes. This is in 
agreement with replica theory. Zero error 
probability can only be achieved in the $N \to \infty $ limit. 
Our simulations are for $N = 10^5 $. It would be interesting to 
perform a detailed study of finite size corrections, i.e. of 
the $N$ dependence of the error probability per bit.\\
We now discuss the numerical value of the noise threshold $w_c^2$. 
The first remark is that both numerically and analytically, 
 the critical value is below Shannon's bound and that 
it depends on the convolutional code (i.e. on the generating 
polynomials). The second remark is that the analytical value of 
thresold, $w_{loc}^2\simeq 1.4186$ db is in very good agreement 
with the numerical value for the 
code \ref{CodiceSemplice}. 
For code \ref{CodiceTipico} $w_{loc}^2 \simeq -2.240$ db while one gets 
$w_c^2 \simeq -1.7$ db from the simulations. 
It would be interesting to understand this disagreement. 
As we said in the previous section, $w_{loc}$ was calculated by 
a local stability analysis of the ordered phase, i.e. we assumed that 
the transition is of second order. A possible explanation would be 
that the transition is second order for code \ref{CodiceSemplice} 
and first order for code 
\ref{CodiceTipico}. Numerical results seem to support this hypothesis, 
as the variation of the error probability as a function of noise 
is much sharper in case \ref{CodiceTipico}. But a much more careful 
analysis of finite size effects is necessary in order to settle this 
question numerically. One should also look analytically for the occurence 
of a first order transition. \\
Another important issue is the breaking of replica symmetry. Since
turbo-decoding is thought to be an approximate algorithm, it may be 
not the best tool to look for replica symmetry breaking. 
We have started an analytical investigation of replica symmetry breaking.

\begin{figure}
\epsfig{figure=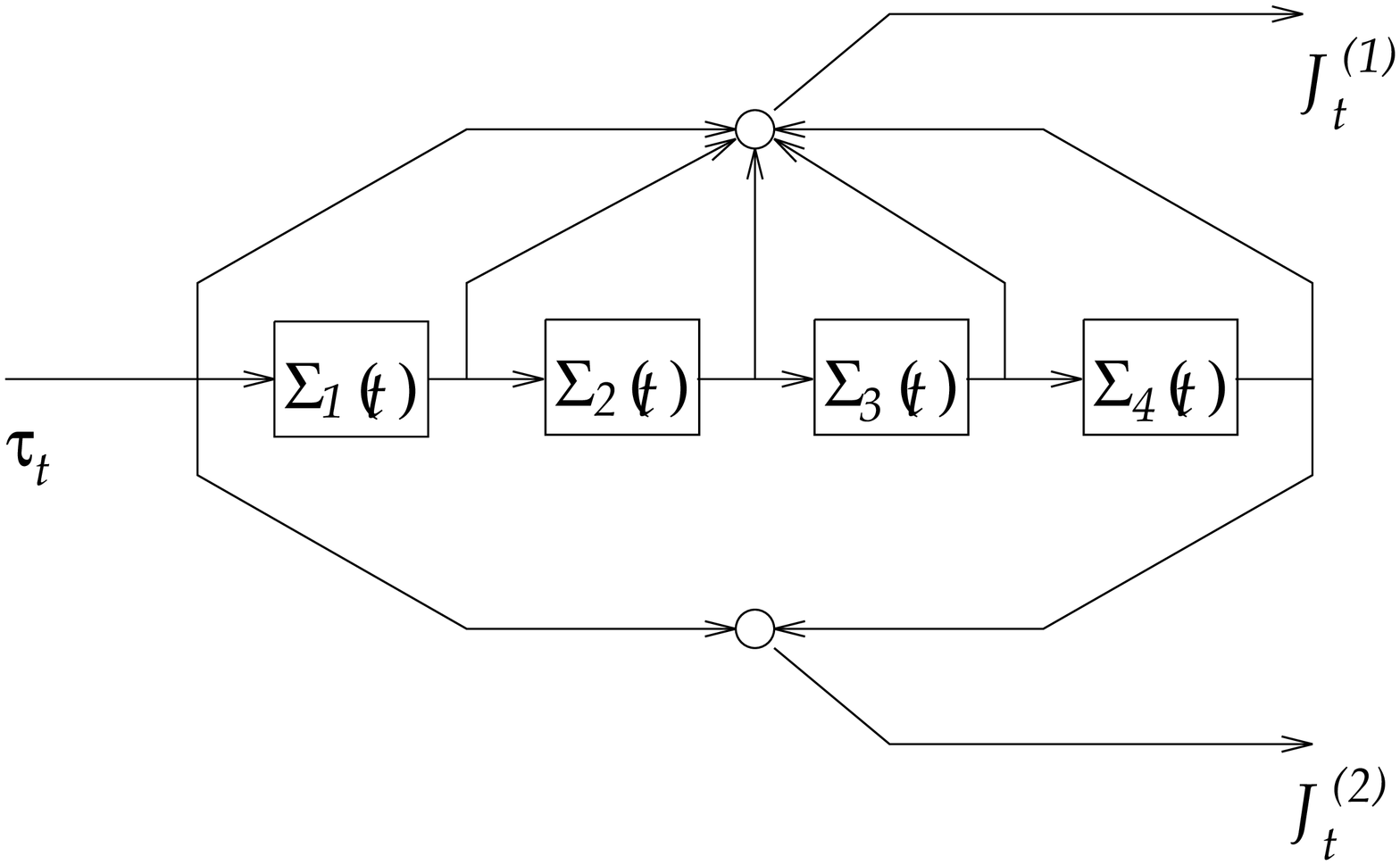,angle=-90,
width=0.85\linewidth}
\caption{Schematic representation of the encoder for a non recursive
convolutional code called code \ref{CodiceComplicato} in the text and
defined by Eqs.(\ref{CodiceComplicato1}),(\ref{CodiceComplicato2}).}
\label{nonrecfig}
\epsfig{figure=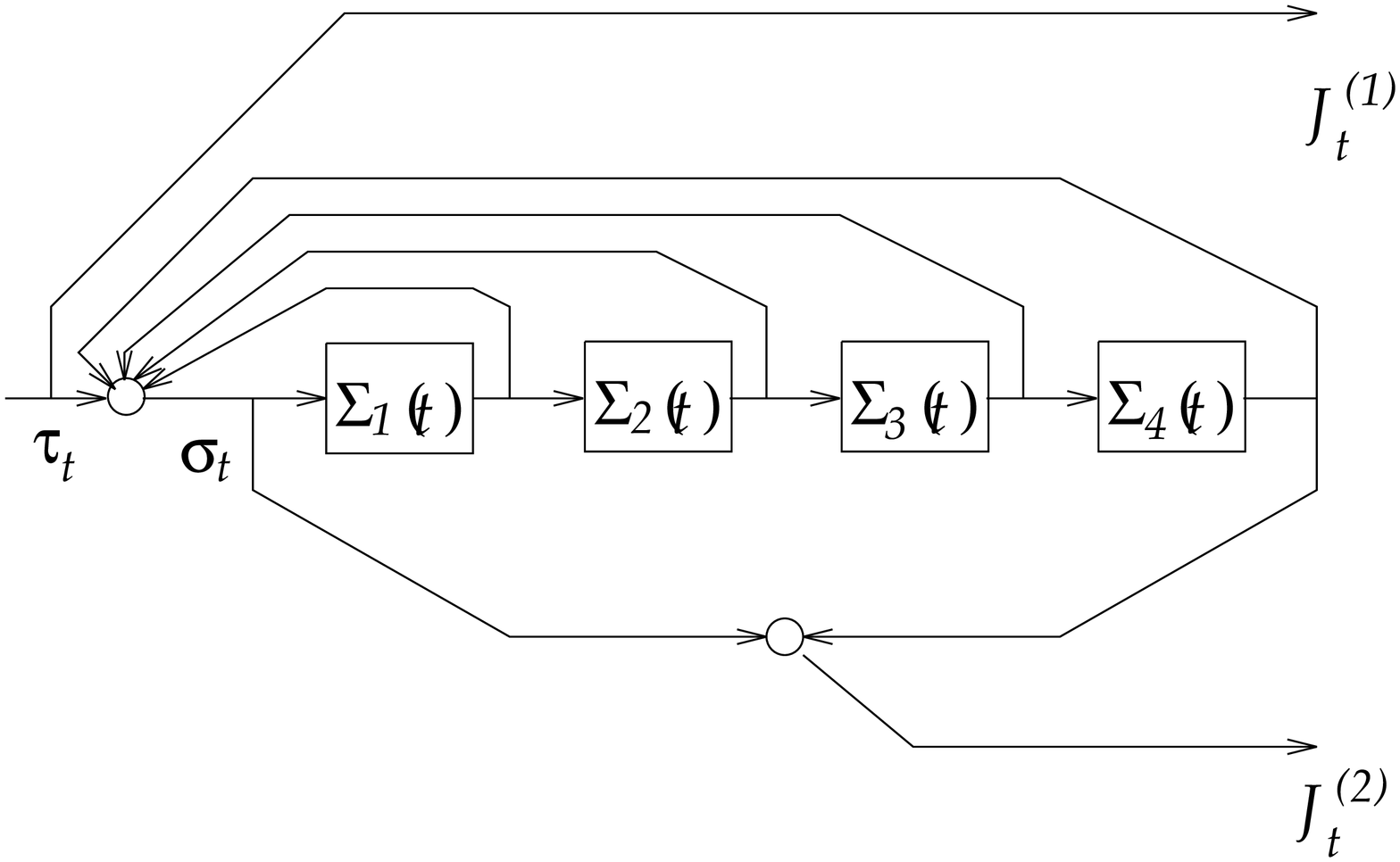,angle=-90,
width=0.85\linewidth}
\caption{The encoder for the recursive convolutional code with the
same generating polynomials as in Fig.(\ref{nonrecfig}) 
(cfr. Eqs.(\ref{CodiceComplicato1}),(\ref{CodiceComplicato2})).}
\label{recfig}
\end{figure}

\begin{figure}
\epsfig{figure=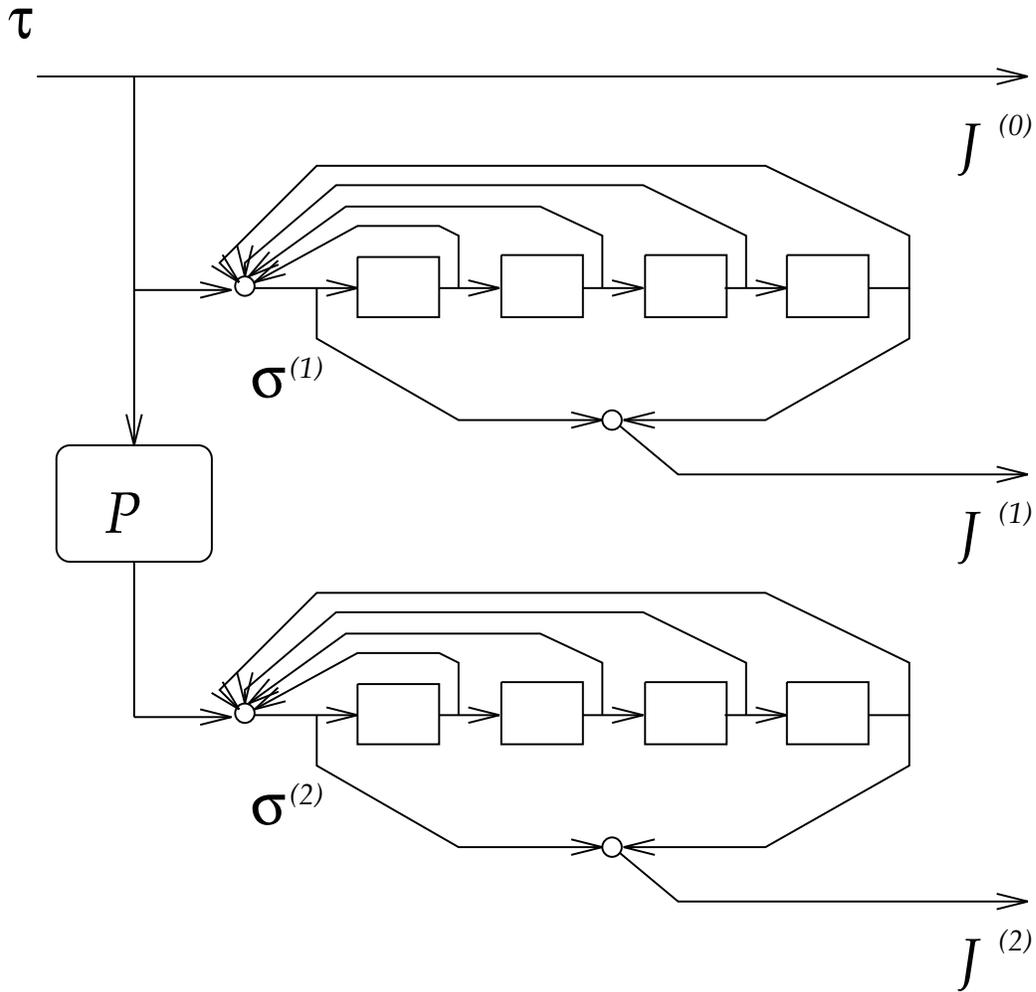,angle=-90,
width=0.85\linewidth}
\caption{Schematic representation of the encoder for a recursive turbo
code with generating polynomials as in the previous figures 
(cfr. Eqs.(\ref{CodiceComplicato1}),(\ref{CodiceComplicato2})).
Notice the presence of the interleaver (denoted by $P$) which
implements the permutation.}
\label{turbofig}
\end{figure}

\clearpage

\begin{thebibliography}{99}

\bibitem{PrimoBerrou} C.Berrou, A.Glavieux, and
P.Thitimajshima. Proc.1993 Int.Conf.Comm. 1064-1070

\bibitem{AlgoritmoViterbi} A.J.Viterbi. IEEE Trans.Com.Technology 
{\bf COM-19}(1971) 751-771

\bibitem{AlgoritmoBCJR} L.Bahl, J.Cocke, F.Jelinek, and J.Raviv. IEEE
Trans.Inf.Theory {\bf IT-20}(1974) 248-287

\bibitem{Sourlas1} N.Sourlas. Nature {\bf 339}(1989) 693-694\\
N.Sourlas, in {\it Statistical Mechanics of Neural Networks}, 
Lecture Notes in Physics 368, ed. L. Garrido, Springer Verlag (1990)\\
N.Sourlas, Ecole Normale Sup{\'e}rieure preprint (April 1993)\\
N.Sourlas, in {\it From Statistical Physics to 
 Statistical Inference and Back,}
 ed. P. Grassberger and J.-P. Nadal, Kluwer Academic (1994), page 195.

\bibitem{DerridaREM} B.Derrida. Phys.Rev. {\bf B 24}(1981) 2613-2626

\bibitem{Rujan} P.Ruj{\'a}n. Phys.Rev.Lett. {\bf 70}(1993) 2968-2971 \\
N.Sourlas. Europhys.Lett. {\bf 25}(1994) 159-164\\
H.Nishimori. J. Phys. Soc. Jpn. {\bf 62}(1993) 2973 

\bibitem{Nishimori} H.Nishimori. J.Phys. {\bf C 13}(1980) 4071-4076 

\bibitem{OccupationDensity} R.Monasson. J.Phys. {\bf A31}(1998) 513-529 

\bibitem{Mio} A.Montanari, in preparation.

\end{thebibliography}
\end{document}